\documentclass[journal]{IEEEtran}
\usepackage[left=1.43cm,right=1.43cm,top=1.8cm,bottom=4.21cm]{geometry}

\usepackage[algo2e]{algorithm2e} 
\usepackage{amsmath,amssymb,amsmath,amsfonts}
\usepackage{bm}
\usepackage{bbm}
\usepackage{dsfont}
\usepackage{graphicx}
\usepackage{cite}
\usepackage{epstopdf}
\usepackage{gensymb}
\usepackage{color}
\usepackage{lipsum}
\usepackage{float}
\usepackage{epstopdf}
\usepackage[mathcal]{eucal}
\usepackage{subfiles}
\usepackage[T1]{fontenc}
\usepackage{siunitx}
\usepackage{tabulary}
\usepackage{epsfig,graphics,graphicx,subfig,amssymb,amstext,amsmath,algorithm,algorithmic,wrapfig,multirow}
\usepackage{array}
\usepackage{adjustbox}
\usepackage[dvipsnames]{xcolor}
\usepackage{cite}
\usepackage{times}
\usepackage{placeins}
\usepackage{textcomp}
\usepackage[font=small]{caption}
\usepackage{flushend}
\usepackage{color, colortbl}
\usepackage{tabularx}
\usepackage{bm}
\usepackage{enumerate}
\usepackage{tikz}
\usepackage{pgfplots}
\usepackage{epstopdf}
\usetikzlibrary{shapes,arrows}
\usepackage[utf8]{inputenc}
\usepackage{mathtools}
\usepackage[utf8]{inputenc}
\usepackage{xcolor}
\usepackage{tikz}
\usetikzlibrary{chains,arrows,calc,positioning}
\usepackage{amssymb}
\usepackage{pifont}
\usepackage{soul}
\usepackage{breqn} 
\usepackage{booktabs} 
\usepackage{multirow}
\usepackage{enumitem}
\usepackage{tabulary}
\usepackage[normalem]{ulem}
\usepackage{dblfloatfix}
\usepackage{makecell}
\usepackage{lipsum}
\usepackage{amsthm}
\usepackage{comment}
\usepackage{filecontents}
\usepackage{bbm}

\newtheoremstyle{mystyle}
  {}
  {}
  {\itshape}
  {}
  {\bfseries}
  {.}
  { }
  {}
\theoremstyle{mystyle}

\newlength \figwidth
\definecolor{bittersweet}{rgb}{1.0, 0.44, 0.37}
\definecolor{glaucous}{rgb}{0.38, 0.51, 0.71}
\definecolor{gainsboro}{rgb}{0.86, 0.86, 0.86}
\definecolor{babyblueeyes}{rgb}{0.63, 0.79, 0.95}
\definecolor{silver}{rgb}{0.75, 0.75, 0.75}
\definecolor{neoncarrot}{rgb}{1.0, 0.64, 0.26}
\definecolor{Gray}{gray}{0.9}
\definecolor{LightCyan}{rgb}{0.88,1,1}
\definecolor{BackgroundLightBlue}{rgb}{0.97,0.97,1}
\definecolor{BackgroundGray}{gray}{0.98}

\makeatletter

 \let\oldforeign@language\foreign@language
 \DeclareRobustCommand{\foreign@language}[1]{%
   \lowercase{\oldforeign@language{#1}}}

\def\nb0{{\mathbf{0}}}
\def\nb1{{\mathbf{1}}}

\def\ncalB{{\mathcal{B}}}

\def\sinr{\mathtt{SINR}}			

\def\calB{\mathcal{B}}

\makeatother

\IEEEoverridecommandlockouts
\begin{document}

\bstctlcite{IEEEexample:BSTcontrol}

\title{Data-Driven Design of 3GPP Handover Parameters with Bayesian Optimization and Transfer Learning}

\author{\IEEEauthorblockN{{Mohamed Benzaghta$^{\star}$, Sahar Ammar$^{\flat}$, David L\'{o}pez-P\'{e}rez}$^{\sharp}$, Basem Shihada$^{\flat}$, and Giovanni Geraci$^{\dagger\,\star}$ \vspace{0.1cm}
} 
\\ \vspace{0.1cm}
\normalsize\IEEEauthorblockA{$^{\star}$\emph{Univ. Pompeu Fabra, Barcelona, Spain} \enspace \enspace $^{\dagger}$\emph{Telefónica Scientific Research, Spain}  \\ $^{\flat}$\emph{King Abdullah Univ. of Science and Technology, Saudi Arabia} \enspace \enspace  $^{\sharp}$\emph{Univ. Politècnica de València, Spain}}

\thanks{This work was supported by 
\emph{a)} HORIZON-SESAR-2023-DES-ER-02 project ANTENNAE (101167288),
\emph{b)} the Spanish Ministry of Economic Affairs and Digital Transformation and the European Union NextGenerationEU through actions CNS2023-145384 and CNS2023-144333, 
\emph{c)} the Spanish State Research Agency through grants PID2021-123999OB-I00 and CEX2021-001195-M, 
\emph{d)} the Generalitat Valenciana, Spain, through the  CIDEGENT PlaGenT, Grant CIDEXG/2022/17, Project iTENTE, and \emph{e)} the UPF-Fractus Chair.}
}

\maketitle
\thispagestyle{empty}
\pagestyle{empty}

\begin{abstract}
Mobility management in dense cellular networks is challenging due to varying user speeds and deployment conditions. Traditional 3GPP handover (HO) schemes, relying on fixed A3-offset and time-to-trigger (TTT) parameters, struggle to balance radio link failures (RLFs) and ping-pongs. We propose a data-driven HO optimization framework based on high-dimensional Bayesian optimization (HD-BO) and enhanced with transfer learning to reduce training time and improve generalization across different user speeds. Evaluations on a real-world deployment show that HD-BO outperforms 3GPP set-1 and set-5 benchmarks, while transfer learning enables rapid adaptation without loss in performance. This highlights the potential of data-driven, site-specific mobility management in large-scale networks.

\end{abstract}


\section{Introduction}
\label{sec:intro}

To meet rising mobile data demands and support new applications, operators are densifying networks to improve coverage, spectral efficiency, and spatial reuse \cite{LopPioGer2025}. However, this densification increases handover (HO) frequency, complicating mobility management. HOs allow user equipment (UE) to maintain quality of service (QoS) while switching between serving and target cells, but frequent transitions pose significant challenges \cite{shahid2024incorporating}.

Given the critical role of mobility management, it has been a focal point for research, industry, and standardization bodies. In \cite{shahid2024incorporating}, the authors propose an HO algorithm designed for small-cell networks that incorporates UE mobility direction to mitigate frequent handovers. Their approach analyzes the reference signal received power (RSRP) pattern during the time-to-trigger (TTT) to determine whether a UE is approaching or moving away from a target cell. Similarly, \cite{chien2024privacy} introduces a federated learning-based handover algorithm that dynamically adjusts HO thresholds based on predicted signal strength and historical performance. The work in \cite{karmakar2022mobility} presents an online learning-based mobility management framework that estimates posterior probabilities of RSRP values to optimize target cell selection. Analytical models have also been explored, such as the mathematical framework in \cite{arshad2021stochastic}, which derives UE-to-base station (BS) association probabilities and HO rates for hybrid radio frequency–visible light communications. Likewise, \cite{malik2024performance} studies HO triggering estimation based on speed, distance, and predefined mobility models.


Mobility management efficiency depends heavily on the configuration of A3-offset and TTT,
which are critical for minimizing radio link failures (RLFs) and ping-pongs.
Handover parameters are typically tuned per cell or within small clusters,
but this localized approach often overlooks inter-cell dependencies and fails to deliver optimal network-wide performance.
While joint optimization could address this, 
it quickly becomes intractable at scale.
Moreover, parameters tuned for specific mobility profiles may not generalize—for instance, high-speed UEs may experience RLFs if TTT delays handover beyond acceptable SINR levels \cite{gures2020comprehensive}.
These challenges highlight the need for cell-specific, data-driven optimization that adapts to local traffic and mobility patterns.

To overcome these challenges, we propose a high-dimensional Bayesian optimization (HD-BO) framework for scalable, robust mobility management across diverse network deployments and UE speeds. To improve convergence and align with 3GPP's vision for model generalization \cite{3GPP38.843}, we also explore transfer learning across varying mobility conditions.

\section{System Model and Problem Formulation}
\label{sec:system_model}
\subsection{Network Topology and Site-Specific Propagation Modeling} 

We consider a real-world radio network deployment operated by a leading commercial mobile provider in the UK. 

\subsubsection*{Network deployment}
The cellular network considered in this study comprises 10 deployment sites, with antenna heights ranging from 22 to 56 meters. Each site hosts three sector antennas, resulting in a total of 30 cells. The deployment covers an urban area of approximately $1400 \times 1275$\,m in London, bounded by latitudes $[51.5087, 51.5215]$ and longitudes $[-0.1483, -0.1296]$. A 3D model of the area is generated using OpenStreetMap data, incorporating both terrain elevation and building structures. Base stations (BSs) are placed and configured to reflect the actual layout and topology of the real-world cellular network.

\subsubsection*{Channel propagation modeling}
The channel between BS $b$ and UE $k$ is modeled using Sionna RT \cite{hoydis2023sionna},
a widely used 3D ray-tracing tool for site-specific radio wave propagation analysis.
Simulations are performed at a carrier frequency of 2 GHz over a 10 MHz bandwidth with 50 physical resource blocks (PRBs).
Buildings are modeled using the \texttt{itu_concrete} material to represent their permittivity and conductivity.
Ray-tracing is configured to allow up to 5 reflections and 1 diffraction, capturing key propagation effects in the urban environment.

\subsubsection*{SINR formulation}
The downlink SINR in dB experienced by UE $k$ from its serving BS $b_k$ is computed as:
\begin{equation}
  \sinr_{\textrm{dB},k} = 10\,\log_{10} \left( \,\frac{p_{b_k} \cdot G_{b_k,k}}{
  \sum\limits_{b\in\calB\backslash b_k}{p_{b} \cdot G_{b,k}  \,+\, \sigma_{\textrm{T}}^2}}\right),
  \label{SINR_DL_TN}
\end{equation}
where $G_{b,k}$ represents the square magnitude of the channel gain, capturing both small-scale and large-scale fading, averaged over the 50 PRBs (each with a bandwidth of 180 kHz). The thermal noise power $\sigma_{\textrm{T}}^2$ over 10\,MHz is obtained from a power spectral density of $-174$\,dBm/Hz, while the transmit power of BS $b$ across the entire bandwidth is $p_{b} = 46$\,dBm \cite{3GPP36814}.

\subsubsection*{Challenges in cellular mobility management}
\begin{figure}
\centering
\includegraphics[width=\figwidth]{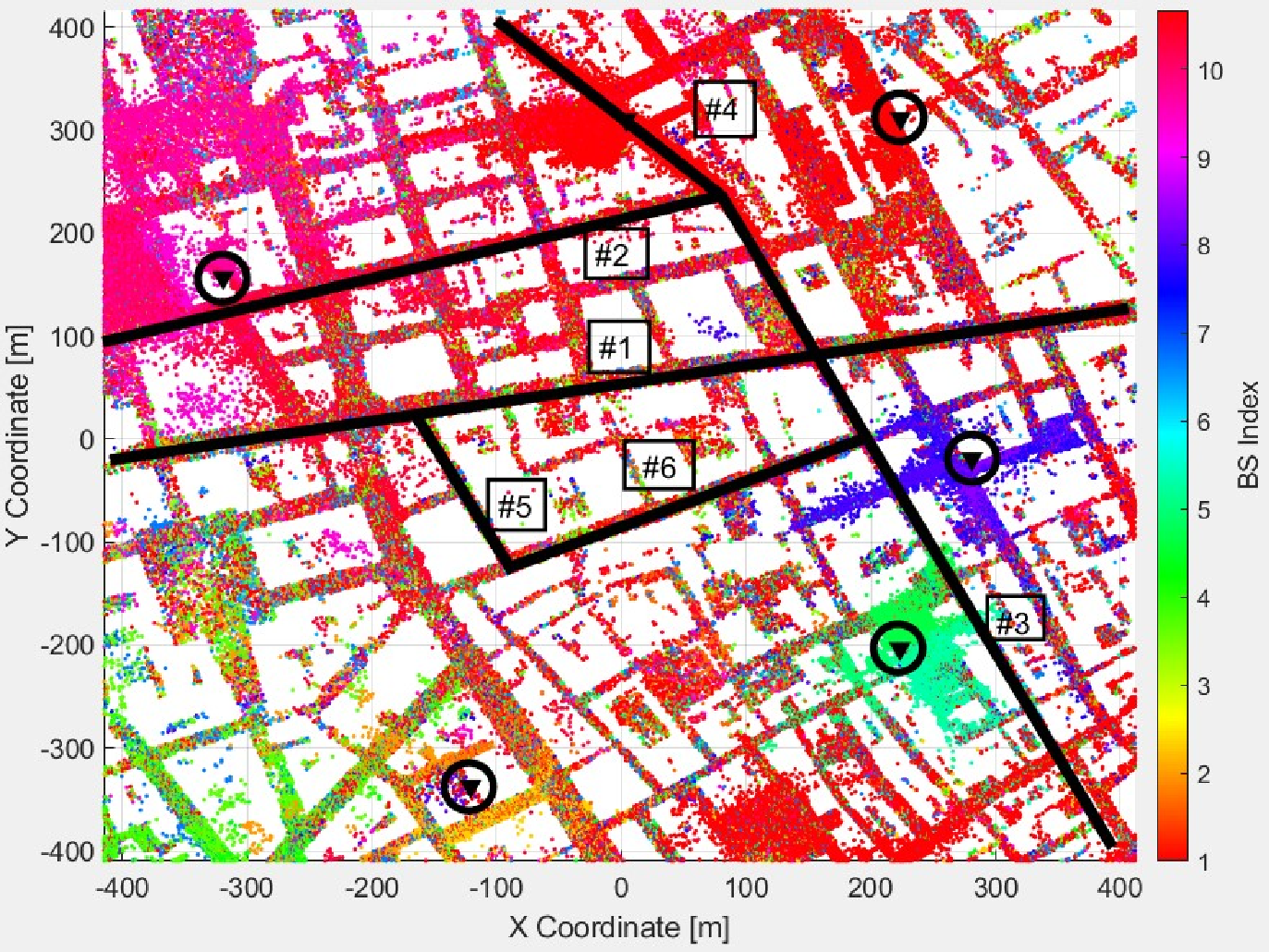}
\caption{2D view of the selected urban area showing UE positions (colored dots) and BS sites (triangles). Black lines indicate the five selected streets; colors represent the cell providing the strongest received power.}
\label{fig:streets}
\end{figure}

We select five main streets in a central London urban area as the focus of our mobility study. Fig.~\ref{fig:streets} presents a 2D map where circled triangles mark BS sites and colored dots represent UE locations along the selected streets, which are highlighted in black and labeled by ID. The dot colors indicate the serving cell site with the highest average received power, as shown in the background heatmap. This spatial distribution illustrates the complexity of real-world mobility management, where rapid signal variations may lead to frequent handovers—highlighting the need for a robust, data-driven framework capable of supporting large-scale deployments and ensuring seamless connectivity.

\subsection{Handover Procedures in Cellular Networks} 

In the following, we provide a detailed overview of the cellular mobility management problem and introduce key concepts, including HO, RLF, and ping-pongs. 

\subsubsection*{Handover process} 
In cellular networks, handovers (HOs) can occur between cells of different radio access technologies (RATs), between cells of the same RAT operating on different frequencies, or between cells of the same RAT and frequency \cite{3GPP36.839}. 
This study focuses on intra-RAT, intra-carrier HOs and specifically on hard HOs, where the UE disconnects from the source cell before establishing a connection with the target cell.

The UE performs HO measurements at both the physical (L1) and network (L3) layers. RSRP values are estimated for all neighboring cells, collected every 40\,ms and averaged over five intervals, a process known as \emph{L1 filtering}. The resulting L1-filtered measurements are further smoothed via a first-order IIR filter at L3---referred to as \emph{L3 filtering}---with a typical period of 200\,ms. This two-stage filtering reduces the impact of fading and measurement noise \cite{lopez2012mobility}. For a visual illustration of the L1 and L3 filtering processes, we refer the reader to Fig.~1 in \cite{lopez2012mobility}.

A HO is triggered when the L3-filtered measurement satisfies a predefined event condition. This study focuses on \emph{Event A3}, which governs intra-RAT, intra-carrier HOs \cite{3GPP36.839, lopez2012mobility}. Event A3 is met when the target cell’s L3-filtered RSRP exceeds that of the serving cell by a hysteresis margin (A3 offset). Once triggered, the UE starts a Time-to-Trigger (TTT) timer. If the condition holds throughout the TTT, the UE sends a measurement report to the serving cell, initiating HO preparation.
 
\subsubsection*{Radio link failure}
Radio link failure (RLF) is triggered when a UE experiences sustained signal degradation. It is governed by the SINR thresholds $Q_{\text{out}}$ and $Q_{\text{in}}$, along with the $T_{310}$ timer. If the downlink SINR ($\sinr_{\textrm{dB},k}$) falls below $Q_{\text{out}}$, a synchronization issue is detected, activating $T_{310}$. If SINR recovers above $Q_{\text{in}}$ before the timer expires, the UE is considered re-synchronized. Otherwise, expiration of $T_{310}$ results in RLF \cite{3GPP36.839}.

\subsubsection*{Handover ping-pongs}
A HO ping-pong occurs when a UE briefly connects to a cell before handing over again. This \emph{time-of-stay} starts with the HO complete message (i.e., indication that the UE has successfully completed the handover) to the target cell and ends with a subsequent HO complete message. A ping-pong is recorded if the time-of-stay is below a threshold $T_p$ (e.g., 1\,s) and the new target cell matches the original source cell. These unnecessary HOs increase signaling overhead and reduce network efficiency.

\subsubsection*{Performance trade-off}
Small A3-offset and TTT values may trigger premature handovers, increasing ping-pongs, while larger values can delay HOs and raise the risk of RLF. Effective mobility management requires tuning these parameters based on UE velocity and site-specific radio conditions. In 3GPP networks, HO decisions are governed by predefined threshold sets. We focus on benchmark configurations set-1 and set-5, which represent the two extremes in HO performance trade-offs. Set-1 delays HOs to reduce ping-pongs (TTT = 480\,ms, A3-offset = 3\,dB), while set-5 accelerates HOs to minimize RLF (TTT = 40\,ms, A3-offset = –1\,dB) \cite{3GPP36.839}. 

\subsection{Problem Formulation} 

Our objective is to determine the optimal configuration of A3-offset and TTT parameters that minimizes conflicting HO key performance indicators (KPIs), specifically the trade-off between RLFs and ping-pongs. The optimization problem is formally defined as follows:

\begin{align}
\min_{\text{\textbf{A3}},\text{\textbf{TTT}}} \;\; w_{\text{PP}} \cdot \frac{\sum_{t} \mathds{1}_{\text{PP}_t}}{\sum_{t} \mathds{1}_{\text{HO}_t} - \mathds{1}_{\text{HOF}_t}} + w_{\text{RLF}} \cdot \frac{\sum_{t} \mathds{1}_{\text{RLF}_t}}{\sum_{t} \mathds{1}_{\text{HO}_t}},
\label{eqn:Opt_problem_joint} 
\end{align}  
%
\begin{align}
\text{s.t.} \quad & \text{A3}_b \in \left( \underline{\text{A3}}, \overline{\text{A3}} \right), \enspace b = 1, \ldots, \ncalB \tag{2a} \\
& \text{TTT}_{b} \in \left( \underline{\text{TTT}}, \overline{\text{TTT}} \right), \enspace b = 1, \ldots, \ncalB \tag{2b}
\end{align}
where $\mathds{1}_{(\cdot)}$ is an indicator function that evaluates whether a handover, RLF, or ping-pong occurs at time step $t$. It returns 1 if the corresponding event is observed, and 0 otherwise. The objective function minimizes a weighted sum of events, with $w_{\text{RLF}}$ and $w_{\text{PP}}$ representing the relative importance of reducing RLF and ping-pongs, respectively. The vectors $\text{\textbf{A3}}$ and $\text{\textbf{TTT}}$ contain the A3-offsets $\text{A3}_b$ and time-to-trigger values $\text{TTT}_b$ for all base stations $b \in \mathcal{B}$. The optimization is subject to bounds, where $\underline{\text{A3}}$ and $\underline{\text{TTT}}$ denote the minimum allowed values, and $\overline{\text{A3}}$, $\overline{\text{TTT}}$ the maximum.

The optimization problem (\ref{eqn:Opt_problem_joint}) is nonconvex due to the nonconcavity of the objective function. Additionally, with 30 cells, the problem involves 60 optimization variables, necessitating efficient techniques to navigate the large search space.
\section{High-dimensional Bayesian Optimization}
\label{sec:BO}
\subsection{Introduction to Bayesian Optimization}

Bayesian Optimization (BO) iteratively builds a probabilistic \textit{surrogate model} of the objective function $f(\cdot)$ using prior evaluations \cite{shahriari2015taking}. This surrogate, which is cheaper to evaluate than $f(\cdot)$, is updated as new data becomes available. An acquisition function $\alpha(\cdot)$ then guides the selection of the next point by balancing exploration (searching for better regions) and exploitation (refining known good solutions).

\subsubsection*{Objective function evaluation}
We define a query point $\textbf{x} = [\textbf{A3}, \textbf{TTT}]$ as the configuration of A3-offset and TTT values for each BS $b \in \mathcal{B}$, and obtain the corresponding objective value $f(\textbf{x})$ from (\ref{eqn:Opt_problem_joint}). The objective function $f(\cdot)$ is a mathematically intractable stochastic function that reflects the system model in Section~\ref{sec:system_model} and accounts for the randomness in UE locations and wireless channels. We evaluate $f(\cdot)$ via system-level simulations, with each evaluation at point $\textbf{x}$ producing a noisy sample $\tilde{f}(\textbf{x})$. In real-world settings, these samples could also be obtained from measurements. Let $\textbf{X} = [\textbf{x}_1,\ldots,\textbf{x}_N]$ denote a set of $N$ query points, and $\textbf{f}(\textbf{X}) = [f_1,\ldots,f_N]^\top$ the corresponding evaluations, where $f_i = f(\textbf{x}_i)$ for $i=1,\ldots,N$.
 
\subsubsection*{Gaussian Process prior distribution}
We use a Gaussian Process (GP) prior, $\widehat{f}(\cdot)$, to construct a surrogate model (i.e., the posterior) that approximates the objective function $f(\cdot)$ \cite{shahriari2015taking}. The resulting GP model enables the prediction of $\tilde{f}(\textbf{x})$ at a query point $\textbf{x}$ based on previously observed values, $\tilde{\textbf{f}}(\textbf{X})=\tilde{\textbf{f}}$, over which the model is trained. 
Formally, the GP prior on the objective function $\tilde{f}(\textbf{x})$ assumes that for any set of input points $\textbf{X}$, the corresponding function values $\tilde{\textbf{f}}$ are jointly distributed as  

\begin{equation}
  p(\,\tilde{\textbf{f}}\,) = \mathcal{N}(\,\tilde{\textbf{f}} \,\,|\,\, \boldsymbol{\mu}(\textbf{X}),\mathbf{K}(\textbf{X})\,),
  \label{posterior}
\end{equation}
where $\boldsymbol{\mu}(\textbf{X}) = [\mu(\mathrm{\textbf{x}}_1),\ldots,\mu(\mathrm{\textbf{x}}_N)]^\top$ is the $N \times 1$ mean vector,  
and $\mathbf{K}(\textbf{X})$ is the $N \times N$ covariance matrix,  
with each entry $(i,j)$ given by the covariance function $k(\textbf{x}_{i},\textbf{x}_{j})$.  
For a given point $\textbf{x}$, the mean function $\mu(\textbf{x})$ provides prior knowledge about $f(\textbf{x})$, while the kernel function $\mathbf{K}(\textbf{X})$ captures the uncertainty between different input values $\textbf{x}$.  

\subsubsection*{Gaussian Process posterior distribution}
Given a set of observed noisy samples $\tilde{\textbf{f}}$ at previously sampled points $\textbf{X}$, the posterior distribution of $\widehat{f}(\textbf{x})$ at a new query point $\textbf{x}$ can be expressed as \cite{frazier2018tutorial}:

\begin{equation}
  p(\widehat{f}(\textbf{x}) = \widehat{f} \,\, | \,\, \textbf{X}, \tilde{\textbf{f}} \,) = \mathcal{N}(\widehat{f} \,\,|\,\, \mu(\textbf{x} \,|\, \textbf{X}, \tilde{\textbf{f}}),\sigma^2(\textbf{x} \,|\, \textbf{X}, \tilde{\textbf{f}})),
  \label{posterior_Noisy}
\end{equation}
where the posterior mean and variance are given by:

\begin{equation}
  \mu(\textbf{x} \,|\, \textbf{X},\tilde{\textbf{f}}) = \mu(\textbf{x}) + \tilde{\textbf{k}}(\textbf{x})^\top (\tilde{\textbf{K}}(\textbf{X}))^{-1}(\tilde{\textbf{f}}-\boldsymbol{\mu}(\textbf{X})),
  \label{Mean_posterior_Noisy}
\end{equation}

\begin{equation}
  \sigma^2(\textbf{x} \,|\, \textbf{X},\tilde{\textbf{f}}) = k(\textbf{x},\textbf{x}) - \tilde{\textbf{k}}(\textbf{x})^\top (\tilde{\textbf{K}}(\textbf{X}))^{-1} \,\tilde{\textbf{k}}(\textbf{x}),
  \label{Kernel_posterior_Noisy}
\end{equation}
where  
$\tilde{\textbf{k}}(\textbf{x}) = [k(\textbf{x},\textbf{x}_{1}),\ldots,k(\textbf{x},\textbf{x}_{N})]^\top$ is the $N \times 1$ covariance vector,  
and $\tilde{\textbf{K}}(\textbf{X}) = \textbf{K}(\textbf{X}) + \sigma^2 \textbf{I}_{N}$,  
with $\sigma^2$ representing the observation noise (i.e., the variance of the Gaussian distribution),  
and $\textbf{I}_{N}$ denoting the $N \times N$ identity matrix.  
Note that \eqref{Mean_posterior_Noisy} and \eqref{Kernel_posterior_Noisy} define the mean and variance of the estimated function $\widehat{f}(\textbf{x})$, where the variance quantifies the uncertainty in the prediction.

\subsubsection*{Initial dataset creation and acquisition function}
The BO algorithm starts by constructing a GP prior $\{\mu(\cdot), k(\cdot, \cdot)\}$ using an initial dataset  
$\mathcal{D} = \{\textbf{x}_1,\ldots,\textbf{x}_{N_{\textrm{o}}},\tilde{f}_1,\ldots,\tilde{f}_{N_{\textrm{o}}}\}$,  
comprising $N_{\textrm{o}}$ initial observations. This dataset is generated via system-level simulations based on the objective in \eqref{eqn:Opt_problem_joint} and the model outlined in Section~\ref{sec:system_model}. For each point $\textbf{x}_i \in \mathcal{D}$, the A3-offset and TTT values are randomly sampled from $[-1\,\text{dB}, 3\,\text{dB}]$ and $[40\,\text{ms}, 480\,\text{ms}]$, respectively. Subsequently, the algorithm selects the next query point $\textbf{x}_n$ using an acquisition function $\alpha(\cdot)$. In this work, we adopt Thompson sampling as the acquisition strategy, due to its effectiveness in navigating exploration–exploitation trade-off \cite{eriksson2019scalable}.
 
\subsubsection*{Limitations of Bayesian Optimization}
While BO \cite{shahriari2015taking} has proven effective for optimizing coverage-capacity tradeoffs and radio resource allocation \cite{benzaghta2023designing, zhang2023bayesian, maggi2023energy, tekgul2023joint}, its scalability is limited to around twenty decision variables in continuous domains \cite{frazier2018tutorial}. This constraint restricts its use for mobility parameter optimization in large-scale cellular networks \cite{de2023towards}. To overcome this, we take the first step toward leveraging High-Dimensional BO (HD-BO) to optimize handover-related KPIs in realistic, large-scale deployments.

\subsection{Introduction to High-dimensional Bayesian Optimization}
To address the scalability limitations of BO in high-dimensional settings, we employ Trust Region Bayesian Optimization (TuRBO) \cite{eriksson2019scalable}.%
\footnote{We implemented and tested three HD-BO methods: Sparse Axis-Aligned Subspaces (SAASBO) \cite[Section~4]{eriksson2021high}, BO via Variable Selection (VSBO) \cite[Section~3]{shen2021computationally}, and Trust Region BO (TuRBO) \cite[Section~2]{eriksson2019scalable}. TuRBO demonstrated superior performance and higher suitability for the problem under consideration.}
Unlike traditional BO, which relies on global surrogate models, TuRBO manages multiple independent local models, each operating within a distinct region of the search space. It achieves global optimization by maintaining several trust regions (TRs) in parallel and allocating samples using an implicit multi-armed bandit strategy, thus focusing exploration on the most promising areas. Each TR employs a GP surrogate model, which retains BO’s key advantages—such as robustness to noise—while avoiding the limitations of overly simple local models. The TR is defined as a hyperrectangle centered at the current best solution $f^*$, with initial side length $L \gets L_{\text{init}}$. The side length of each dimension is then scaled according to its corresponding GP length scale $\lambda_i$, ensuring adaptive resolution across dimensions. The side length for each dimension is given by:
\begin{equation}
L_i = {\lambda_i L}\cdot{\left(\sideset{}{_{j=1}^d}\prod \lambda_j \right)^{-1/d}}\!\!\!\!\!\!.
\end{equation}
where $d$ is the total number of dimensions (i.e, optimization parameters under consideration). During each local optimization run, an acquisition function selects a batch of $q$ candidates at each iteration, ensuring they remain within the designated TR. If the TR’s side length $L$ were large enough to cover the entire search space, this method would be equivalent to standard vanilla-BO. Thus, adjusting $L$ is crucial: the TR must be large enough to encompass promising solutions while remaining compact enough to ensure the local model's accuracy.  
The TR is dynamically resized based on optimization progress: it is doubled ($L \gets \min \{L_{\text{max}}, 2L\}$) after $\tau_{\text{succ}}$ consecutive successes and halved ($L \gets L/2$) after $\tau_{\text{fail}}$ consecutive failures. Success and failure counters are reset after each adjustment. If $L$ falls below $L_{\text{min}}$, the TR is discarded, and a new one is initialized at $L_{\text{init}}$. The TR’s side length is capped at $L_{\text{max}}$.  
TuRBO maintains $m$ trust regions simultaneously, denoted as $\text{TR}_{l}$, where $l \in \{1, \dots, m\}$, each defined as a hyperrectangle with a base side length $L_{l} \leq L_{\text{max}}$. Candidate selection involves choosing a batch of $q$ candidates from the union of all TRs. Thompson sampling is used for selecting candidates both within and across trust regions.  

In this study, TuRBO runs using an open-source repository \cite{eriksson2019scalable} with the following hyperparameters: $\tau_{\text{succ}} = 3, \tau_{\text{fail}} = 15, L_{\text{init}} = 0.8, L_{\text{min}} = 2^{-7}, L_{\text{max}} = 1.6$.

\section{Numerical Results}
To evaluate the HD-BO framework, we jointly optimize A3-offset and TTT to minimize the objective in~\eqref{eqn:Opt_problem_joint}. Our case study examines UE mobility at 3, 30, and 60\,km/h across five main streets in London. We also assess HD-BO’s ability to generalize across user speeds, focusing on its performance with transfer learning. For our evaluations, we set $Q_{\text{out}} = -8$ dB,  $T_{310}$ = 1s, and $T_p$ = 1s, as per \cite{3GPP36.839}.

\subsection{Performance of Data-driven Mobility Management}

\begin{figure}
\centering
\includegraphics[width=\figwidth]{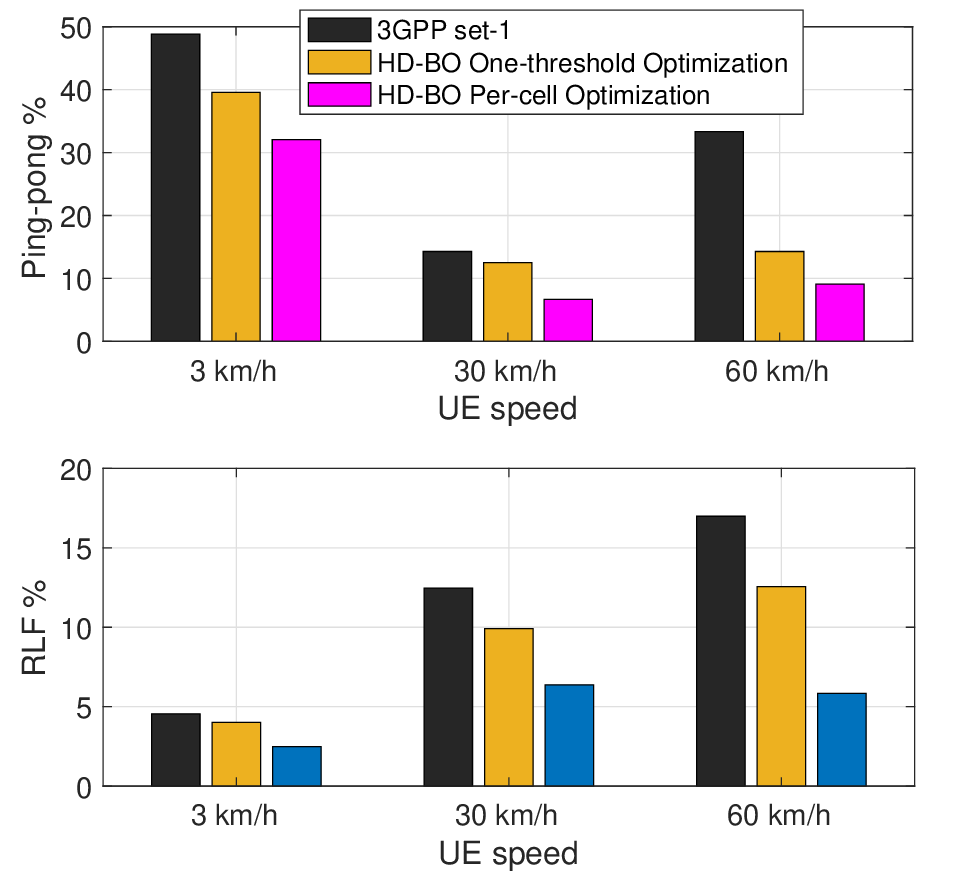}
\caption{RLFs and ping-pongs at different speeds under HD-BO (one-threshold and per-cell) and 3GPP baseline. Lower is better.}
\label{fig:allcells_vs_percell_pingpongs_RLF_compare}
\end{figure}

\begin{figure}
\centering
\includegraphics[width=\figwidth]{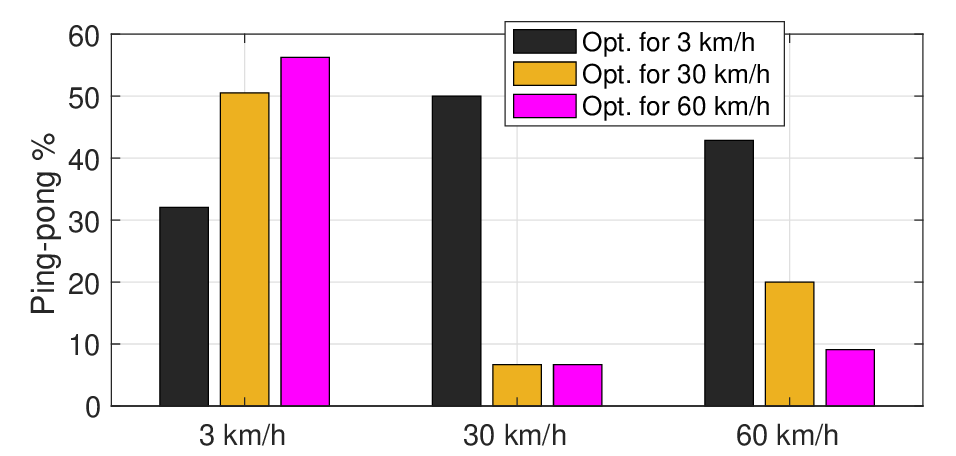}
\caption{Ping-pongs for UEs at different speeds with HD-BO optimized for a single target speed. Lower is better.}
\label{fig:different_speeds_opt_pingpongs_compare}
\end{figure}

Fig.~\ref{fig:allcells_vs_percell_pingpongs_RLF_compare} shows RLF and ping-pong performance for UEs at 3, 30, and 60\,km/h under HD-BO, comparing one-threshold (all cells share the same configuration) and per-cell optimization with $w_{\text{PP}} = 9$ and $w_{\text{RLF}} = 1$. Fig.~\ref{fig:different_speeds_opt_pingpongs_compare} illustrates how optimizing for one speed affects others. Key observations are as follow:

\begin{itemize}
\item
Compared to the 3GPP set-1 benchmark (tuned for ping-pong reduction), HD-BO with one-threshold optimization reduces ping-pongs by 20\%. Per-cell optimization yields further gains, achieving a 35\% reduction at 3\,km/h. At 60\,km/h, improvements are more significant—57\% and 73\% reductions for one-threshold and per-cell cases, respectively (see Fig.~\ref{fig:allcells_vs_percell_pingpongs_RLF_compare}). These results highlight the advantage of per-cell optimization in mobility management.

\item 
Optimizing handover parameters for a specific speed using HD-BO may degrade performance at other speeds. For example, a configuration optimized for 3\,km/h leads to an 8$\times$ increase in ping-pongs at 30\,km/h (Fig.~\ref{fig:different_speeds_opt_pingpongs_compare}), as higher speeds require shorter TTT values to ensure timely handovers. These results highlight the need to apply HD-BO across all speed categories—3, 30, and 60\,km/h—to achieve robust and balanced mobility performance.
\end{itemize}

\begin{figure}
\centering
\includegraphics[width=\figwidth]{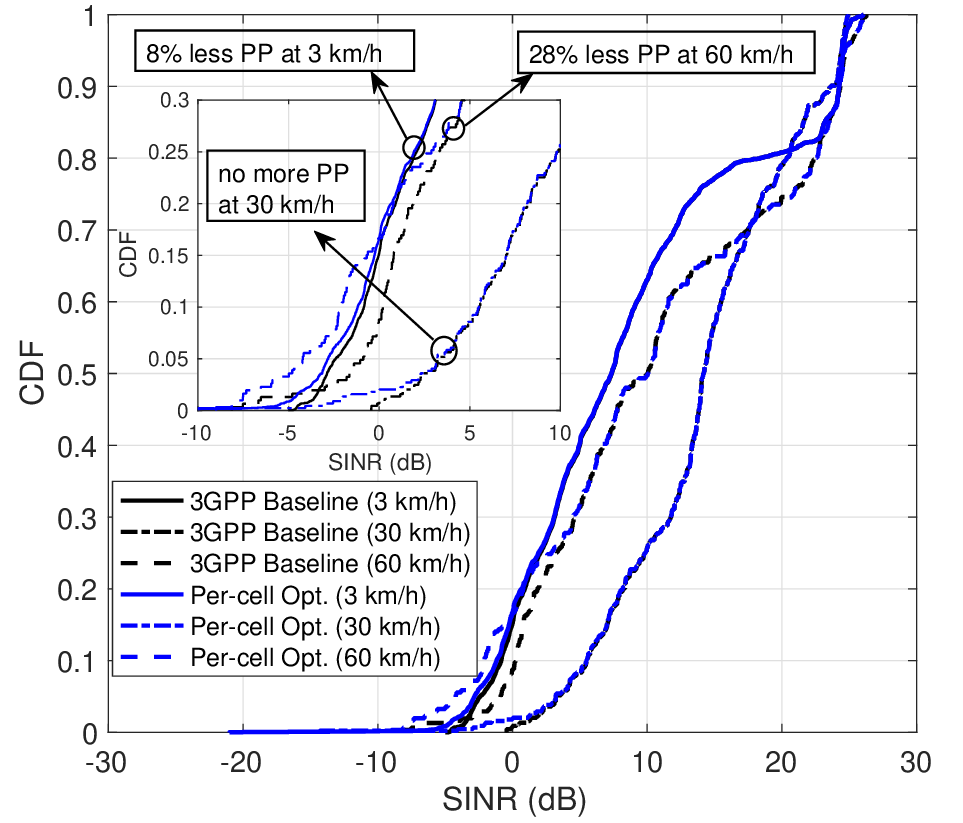}
\caption{SINR for UEs at various speeds under HD-BO ($w_{\text{PP}} = 1$, $w_{\text{RLF}} = 9$) and 3GPP baseline.}
\label{fig:SINR_GUE_mixSpeed}
\end{figure}

Fig.~\ref{fig:SINR_GUE_mixSpeed} shows the CDF of downlink SINR for UEs at different speeds when the network is jointly optimized across all speed categories using HD-BO. The black curves denote the 3GPP set-5 benchmark (upper bound), which enables immediate HOs without considering ping-pongs. The blue curves correspond to HD-BO optimization with $w_{\text{PP}} = 1$ and $w_{\text{RLF}} = 9$, prioritizing RLF reduction across five streets. HD-BO achieves SINR performance close to the 3GPP upper bound while reducing ping-pongs by 8\% (3\,km/h), 28\% (60\,km/h), and 100\% (30\,km/h).

\subsection{Transfer Learning with HD-BO}

For commercial deployment, machine learning models must perform reliably across diverse operating conditions \cite{lin2023overview}. This subsection evaluates the generalization capability of HD-BO across different UE speeds via transfer learning. In optimization, transfer learning accelerates convergence on a new task (target) by reusing data or knowledge from a related task (source), particularly useful when generating the initial BO dataset $\mathcal{D}$ is costly (e.g., real-world measurements). Let $\mathcal{D}_{\text{sr}}$ and $\mathcal{D}_{\text{tg}}$ denote datasets for the source and target scenarios, respectively. We compare three configurations: 100\% target data ($\mathcal{D} = \mathcal{D}_{\text{tg}}$), 100\% source data ($\mathcal{D} = \mathcal{D}_{\text{sr}}$), and a 50\%/50\% mix. The source scenario features UEs moving at 60\,km/h, while the target scenario uses the same deployment but at 30\,km/h. The objective is to assess whether HD-BO can adapt to new mobility patterns without full retraining.

\subsubsection*{Convergence and performance of transfer learning}
\begin{figure}
\centering
\includegraphics[width=\figwidth]{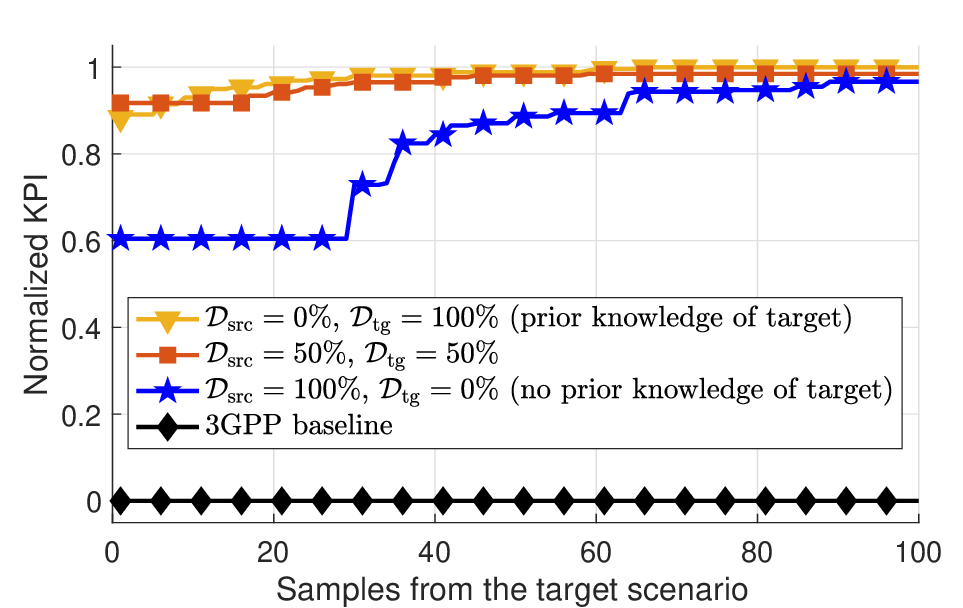}
\caption{Convergence of transfer learning with diverse UE speeds.}
\label{fig:LT_convg}
\end{figure}

Fig.~\ref{fig:LT_convg} illustrates the convergence behavior of transfer learning using HD-BO by tracking the best observed objective at each iteration $n$. To provide a practical performance measure, we plot the min-max normalized KPI, where 0 represents the 3GPP baseline (worst performance), and 1 corresponds to the upper bound—the best achievable KPI when the HD-BO posterior is fully trained on the target scenario ($\mathcal{D} = \mathcal{D}_{\text{tg}}$). The initial dataset $\mathcal{D}$ consists of $N_{\textrm{o}} = 60$ observations.  

\begin{figure}
\centering
\includegraphics[width=\figwidth]{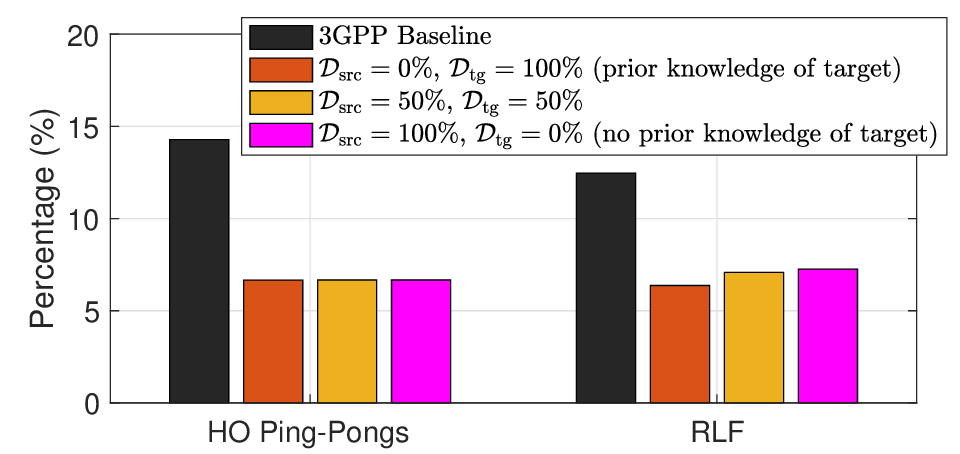}
\caption{Ping-pongs and RLF with transfer learning. Lower is better.}
\label{fig:LT_perform}
\end{figure}

Fig.~\ref{fig:LT_convg} shows that when 50\% of $\mathcal{D}$ is sourced from the target scenario, convergence occurs at a rate comparable to the case where 100\% of $\mathcal{D}$ is from the target scenario, demonstrating that transfer learning can effectively reduce the need for extensive new data collection.
Fig.~\ref{fig:LT_perform} compares the achieved RLF and ping-pong percentages across different initial dataset configurations. The results indicate comparable performance among all three dataset variations. Even without target-specific prior knowledge ($\mathcal{D} = \mathcal{D}_{\text{sr}}$), performance degrades by only 3\%, underscoring the ability to generalize across related mobility scenarios.
 
\section{Conclusion }
\label{sec:conclusion}
We proposed a data-driven mobility management framework based on high-dimensional Bayesian optimization (HD-BO) for optimizing handovers in cellular networks. By jointly optimizing the A3-offset and TTT parameters across all cells, our framework effectively balances the trade-off between ping-pongs and radio link failures in site-specific scenarios. Moreover, since we leverage optimizing 3GPP standard-compliant parameters, our approach is readily applicable to real-world networks. We also demonstrated that transfer learning improves model generalization, reducing training overhead while maintaining performance. Practical evaluations across diverse UE speeds confirm the framework’s effectiveness in dense urban deployments. Future directions include extending the method to beam-based mobility in mmWave systems and supporting vertical handovers in integrated terrestrial and non-terrestrial networks \cite{geraci2022integrating,benzaghta2022uav}.

\bibliographystyle{IEEEtran}
\bibliography{journalAbbreviations, main}

\end{document}